\documentclass{PoS}

\usepackage{longtable}
\usepackage{booktabs}
\usepackage{caption}

\title{IceCube as a Multi-messenger Follow-up Observatory for Astrophysical Transients}

\ShortTitle{IceCube Fast Response Pipeline}

\author{
The IceCube Collaboration\footnote{For collaboration list, see PoS(ICRC2019) 1177.}\\
{\itshape \href{http://icecube.wisc.edu/collaboration/authors/icrc19_icecube}{http://icecube.wisc.edu/collaboration/authors/icrc19\_icecube}}\\
E-mail: \email{justin.vandenbroucke@wisc.edu}
}

\abstract{

The recent association between IC-170922A and the blazar TXS0506+056 highlights the importance of real-time observations for identifying possible astrophysical neutrino sources.
Thanks to its near-100\% duty cycle, 4$\pi$ steradian field of view, and excellent sensitivity over many decades of energy, IceCube is well suited both to generate alerts for follow-up by other instruments and to rapidly follow up alerts generated by other instruments. Detection of neutrinos in coincidence with transient astrophysical phenomena serves as a smoking gun for hadronic processes and supplies essential information about the identities and mechanisms of cosmic-ray accelerators. In 2016, the IceCube Neutrino Observatory established a pipeline to rapidly search for neutrinos from astrophysical transients on timescales ranging from a fraction of a second to multiple weeks. Since then, 67 dedicated analyses have been performed searching for associations between IceCube neutrinos and astrophysical transients reported by radio, optical, X-ray, and gamma-ray instruments in addition to searching for lower energy neutrino signals in association with IceCube's own high-energy alerts. We present the event selection, maximum likelihood analysis method, and sensitivity of the IceCube real-time pipeline. We also summarize the results of all follow-up analyses to date. \\

\vspace{4mm}
{\bfseries Corresponding authors:}
Kevin Meagher$^{1}$, Alex Pizzuto$^{1}$, \speaker{Justin Vandenbroucke}$^{1}$\\
{$^{1}$ \itshape Department of Physics and Wisconsin IceCube Particle Astrophysics Center, University of Wisconsin, Madison, WI 53706, USA}\\
}

\FullConference{36th International Cosmic Ray Conference -ICRC2019-\\
		July 24th - August 1st, 2019\\
		Madison, WI, U.S.A.}

\begin{document}

\section{Introduction}\label{sec:intro}

Major breakthroughs in multi-messenger astronomy have been made recently thanks to real-time alerts generated by observatories and rapidly followed up by other instruments.  This includes the first electromagnetic counterparts of a gravitational wave source~\cite{GWMM} and evidence for neutrino emission by a flaring blazar~\cite{TXSIceCube,TXSMM}.  The latter was possible thanks to a real-time neutrino alert generated by the IceCube Neutrino Observatory and followed up by two dozen observatories including (in gamma rays) the  Fermi Large Area Telescope, MAGIC, and VERITAS. Because IceCube views the entire sky (both the Northern hemisphere and, with reduced sensitivity, the Southern hemisphere) with over 99\% duty cycle, there is excellent potential for additional discoveries in the opposite direction: IceCube can search for neutrinos from arbitrary directions and times in response to interesting astrophysical phenomena detected by other observatories.

\begin{figure}[h!]
  \centering
    \includegraphics[width=4in]{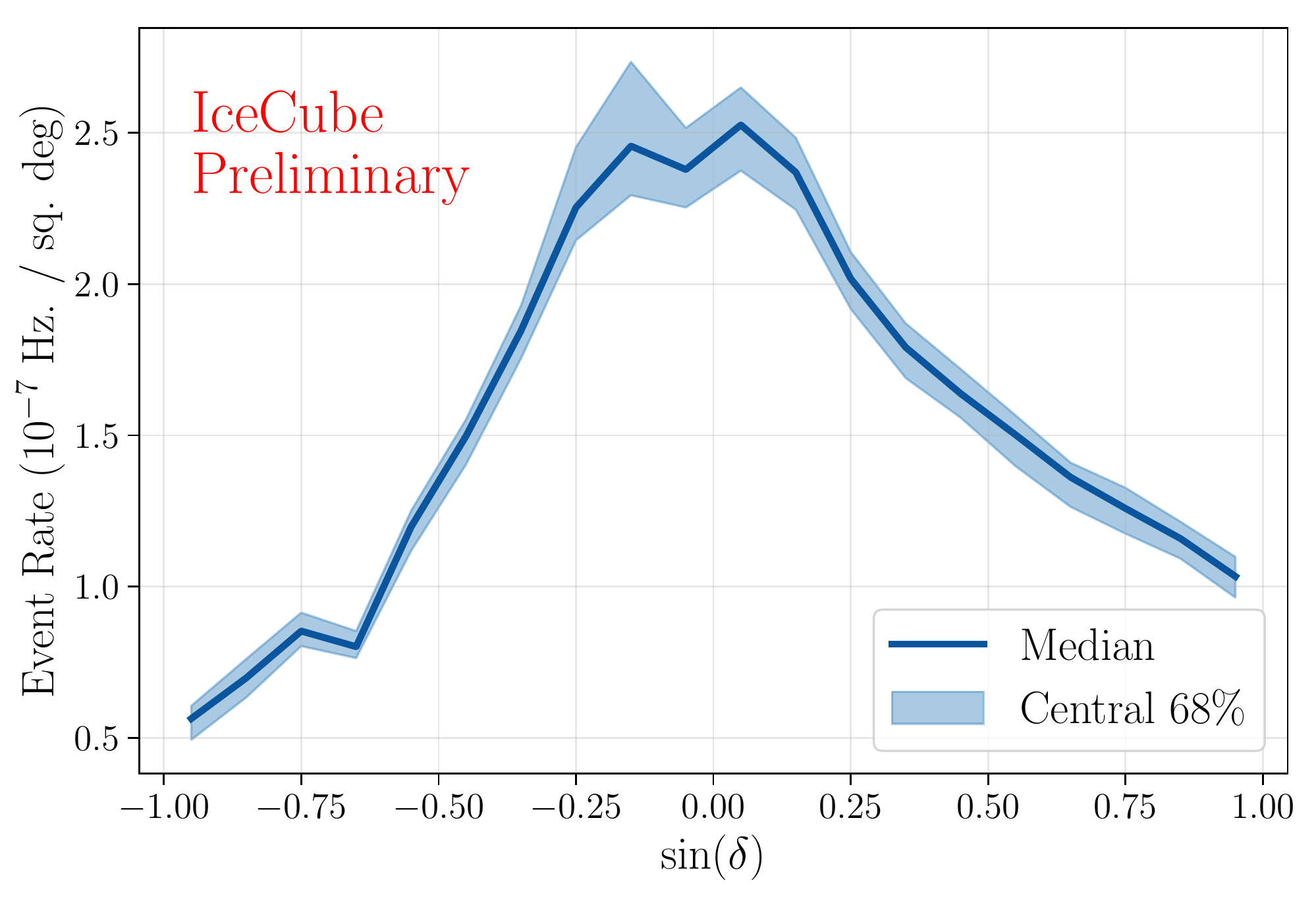}
    \caption{Rate of events per square degree in the low-latency online event selection.  The range of variation due to seasonal modulation of the atmospheric muon and neutrino rates is indicated by the shaded band.\label{rate}}
\end{figure}

Astrophysical transients including blazar flares, supernovae, tidal disruption events, fast radio bursts, microquasar flares, and gravitational wave sources including gamma ray bursts are all possible high-energy neutrino sources. IceCube developed and executes a rapid response pipeline in order to search for neutrinos from these or any other astrophysical objects or multi-messenger events, given particular directions and time windows determined from observations by other instruments. The pipeline uses a selection of muon neutrino candidate track events called the ``gamma-ray followup'' (GFU) event selection because it was originally developed for triggering rapid very-high-energy gamma-ray observations~\cite{GFU}.  The events are identified quickly at the South Pole and relayed over satellite to the University of Wisconsin with 0.5~minute mean latency. The all-sky average GFU event rate is $\sim$6~mHz.  This background rate has modest ($\sim$10\% peak-to-peak) annual modulation due to seasonal variation of the atmospheric temperature and pressure.  The event rate as a function of declination is shown in Figure~\ref{rate}.  Events in the Northern hemisphere are predominantly atmospheric neutrinos and those in the Southern hemisphere are predominantly atmospheric muons.

In addition to well localized sources (with uncertainty smaller than the $\sim$1$^\circ$ angular reconstruction uncertainty of events in the GFU sample), the pipeline can search for (1) neutrino emission with a radially symmetric, extended spatial distribution, or (2) point-like neutrino emission within an extended search region of arbitrary probability distribution on the sky.  This is the same feature of the analysis framework that is used for IceCube's rapid follow-up of gravitational wave events~\cite{GW_ICRC}, and it is also useful for other sources with extended localization regions including fast radio bursts detected by radio interferometers~\cite{FRB} and tau neutrino candidates detected by the ANITA neutrino detector~\cite{ANITA}.

In addition to executing the pipeline in response to external triggers, we run it in response to IceCube's own high-energy neutrino alerts to search for accompanying events at lower energy in the same direction. For each high-energy neutrino alert, two time windows are searched: (-1, +1)~day and (-30, +1)~day with respect to the alert neutrino time. For these analyses, the IceCube event that triggered the alert is excluded from the analysis sample.  The two durations are selected to strike a balance between model independence and background control: larger signal search windows are sensitive to all signal durations up to the search window, but with increased background.

\begin{figure}[h!]
  \centering
    \includegraphics[width=2.9in]{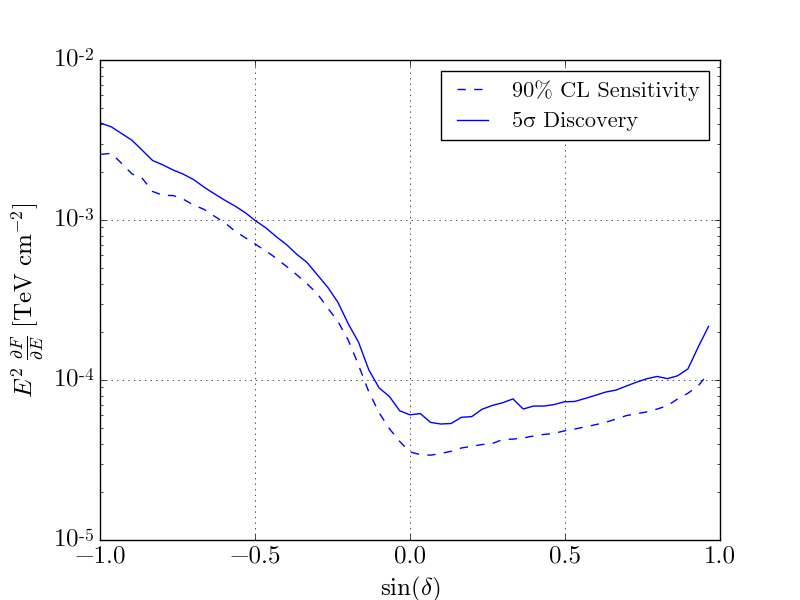}
    \includegraphics[width=2.9in]{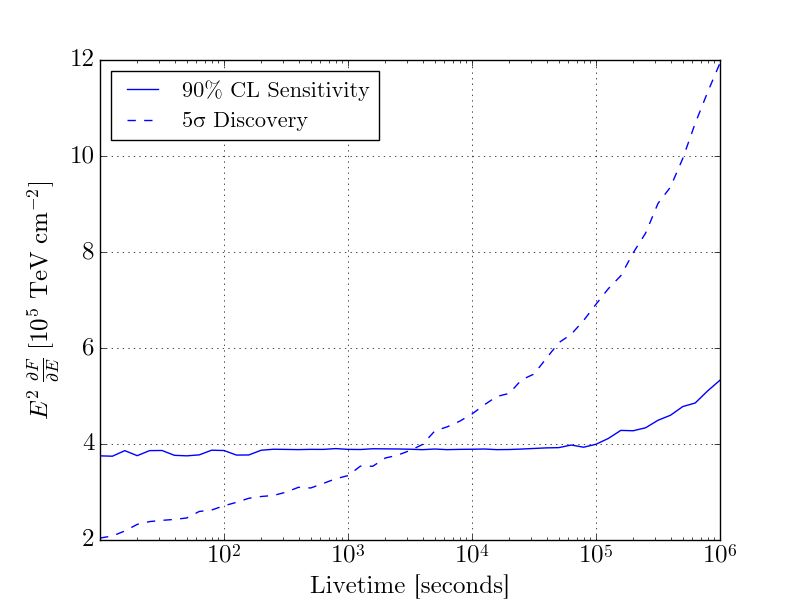}
    \caption{(Left) Pipeline performance as a function of declination, for a 10$^5$~s analysis time window and zero spatial source extension. Most of the declination dependence is a result of background rate variation. The solid line shows the discovery potential, i.e., the minimum flux required to achieve a 5$\sigma$ discovery in 50\% of signal plus background realizations (including Poisson fluctuations of each).  The dashed line shows sensitivity, i.e., the 90\% confidence level upper limit corresponding to the the median value of the test statistic for the background-only hypothesis.  (Right) Pipeline sensitivity and discovery potential as a function of signal window duration, for declination $\delta = 15^\circ$ and zero spatial source extension. Since the background rate is low, the sensitivity remains flat for timescales less than 10$^5$ seconds. In both plots the flux is time integrated over the analysis window duration.  \label{sensitivity}}
\end{figure}

\section{Method and sensitivity}\label{sec:method}

The fast response pipeline uses the same unbinned maximum likelihood method (described in ~\cite{BraunTimeIndep, BraunTimeDep}) used for most IceCube searches for point-like astrophysical neutrino sources.  In particular, because this is a search for transient or flaring emission, the likelihood includes a Poisson term that compares the total number of events observed in the search window with the number expected from background.  This is the same method that is used for IceCube analyses of gamma-ray bursts~\cite{GRB} and fast radio bursts~\cite{FRBJournal}.  This is in contrast to searches for steady neutrino emission, such as~\cite{PointSource}, in which comparing the total observed counts with the total expected counts is not very helpful because such analyses are highly background dominated.  The likelihood includes an energy term in order to weight higher energy events as more signal-like than lower energy events under the expectation that astrophysical neutrino sources have harder spectra than the atmospheric muon and neutrino spectra.  This expectation is supported by measurements of the diffuse astrophysical neutrino spectrum.  The analysis is optimized for an $E^{-2}$ power law spectrum and is also sensitive to other similar spectra.

The fast response pipeline method and performance were reported in ~\cite{icrc2017}.  Figure~\ref{sensitivity} shows the sensitivity and discovery potential of the pipeline as a function of declination and signal time window duration.  For duration below $\sim$10$^5$ seconds, the sensitivity (average upper limit that is set in the absence of a detection) is independent of duration because the expected number of background events is well below 1.  The number of events required to achieve 5~$\sigma$ detection depends on the source declination and analysis time window as well as the energy and angular uncertainty of the detected events.  For some analysis configurations with short search time window, one event is sufficient for 5~$\sigma$ detection.

% TODO: do we include transient Poisson term in likelihood?
%  TODO: add GFU latency plot?

\begin{figure}[h!]
  \centering
    \includegraphics[width=4in]{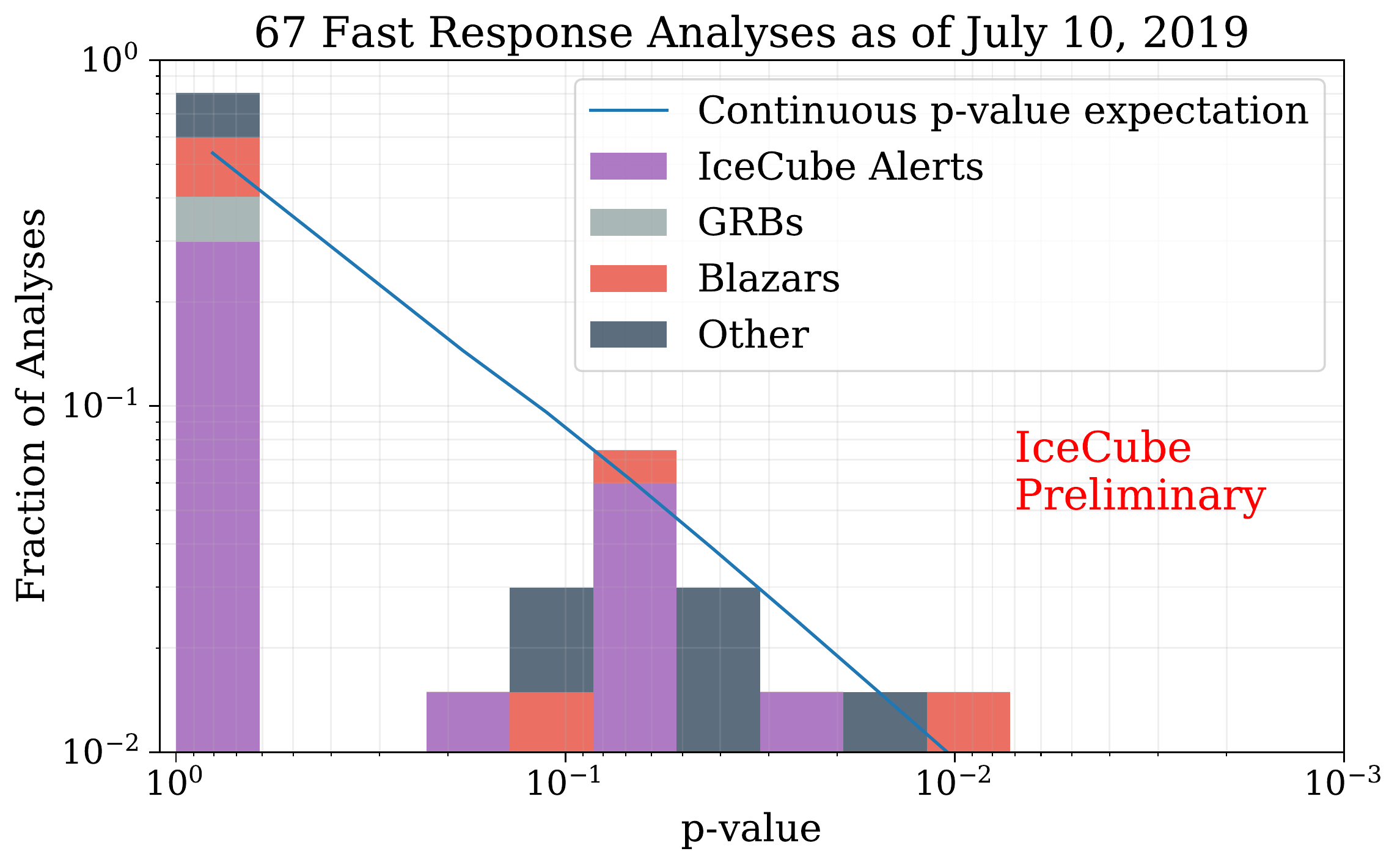}
    \caption{P value distribution for all analyses run by the fast response pipeline, separated according to source class, as of July 10, 2019.  No analysis has resulted in a statistically significant signal detection.  Analyses typically have a short search time window, with an expected number of background events significantly less than one.  In this regime, the unbinned maximum likelihood method performs approximately the same as a binned Poisson counting experiment whose p value is determined by the number of counts.  Because of this, the p value distribution under the background-only null hypothesis is discrete, with a high fraction of occurrences at $p=1.0$ corresponding to zero events in the signal window, and a separate cluster of p values corresponding to one event in the signal window.  The expectation for a continuous, uniform p value distribution is shown for reference.\label{pvalues}}
\end{figure}

\section{Results and outlook}\label{sec:results}

As of July 10, 2019, the pipeline has been run 67 times.  The target object and results for each analysis are summarized in Table~\ref{table}.  The distribution of p values from individual analyses is shown in Figure~\ref{pvalues}.

\begin{figure}[h]
  \centering
    \includegraphics[width=2.9in]{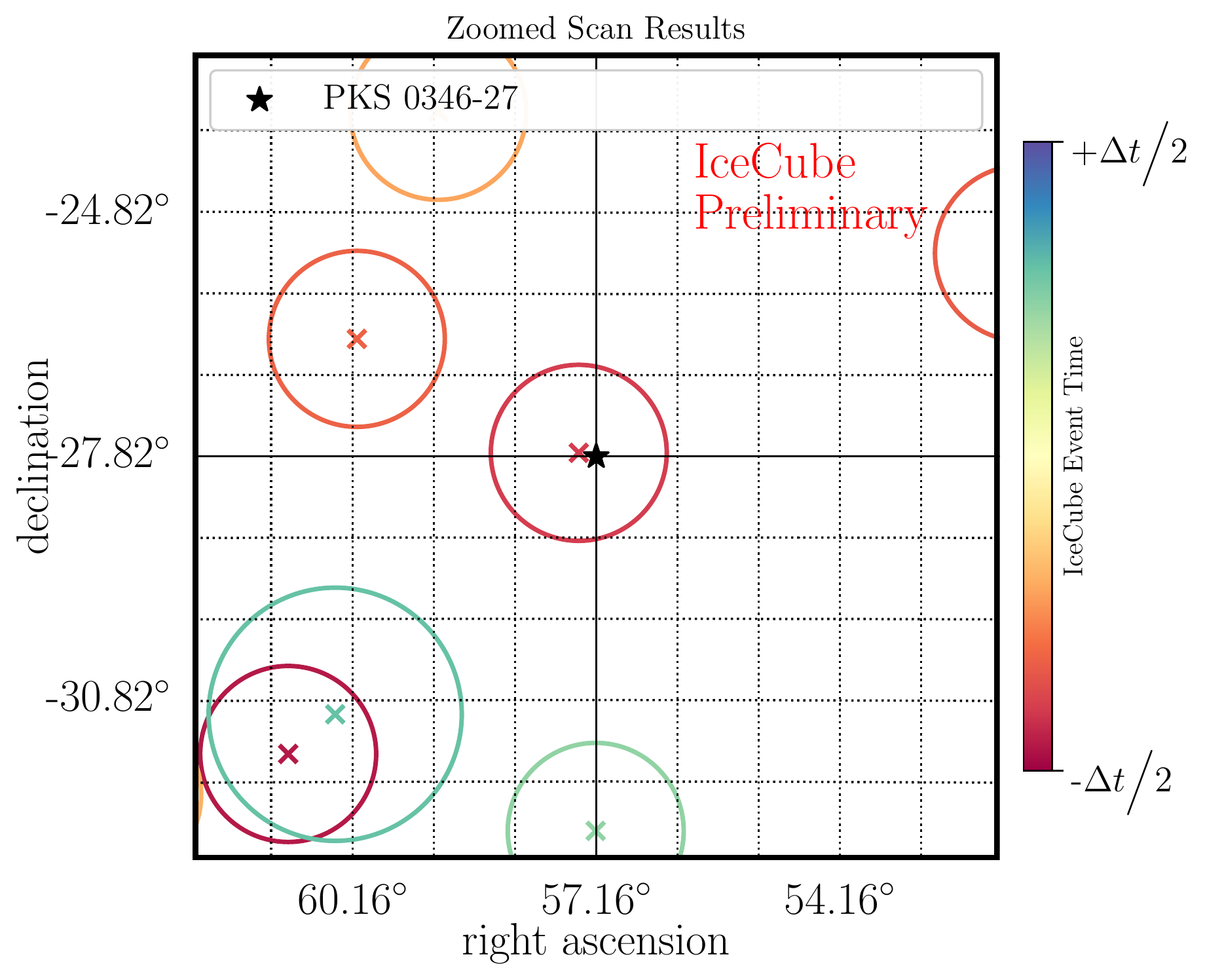}
    \includegraphics[width=2.9in]{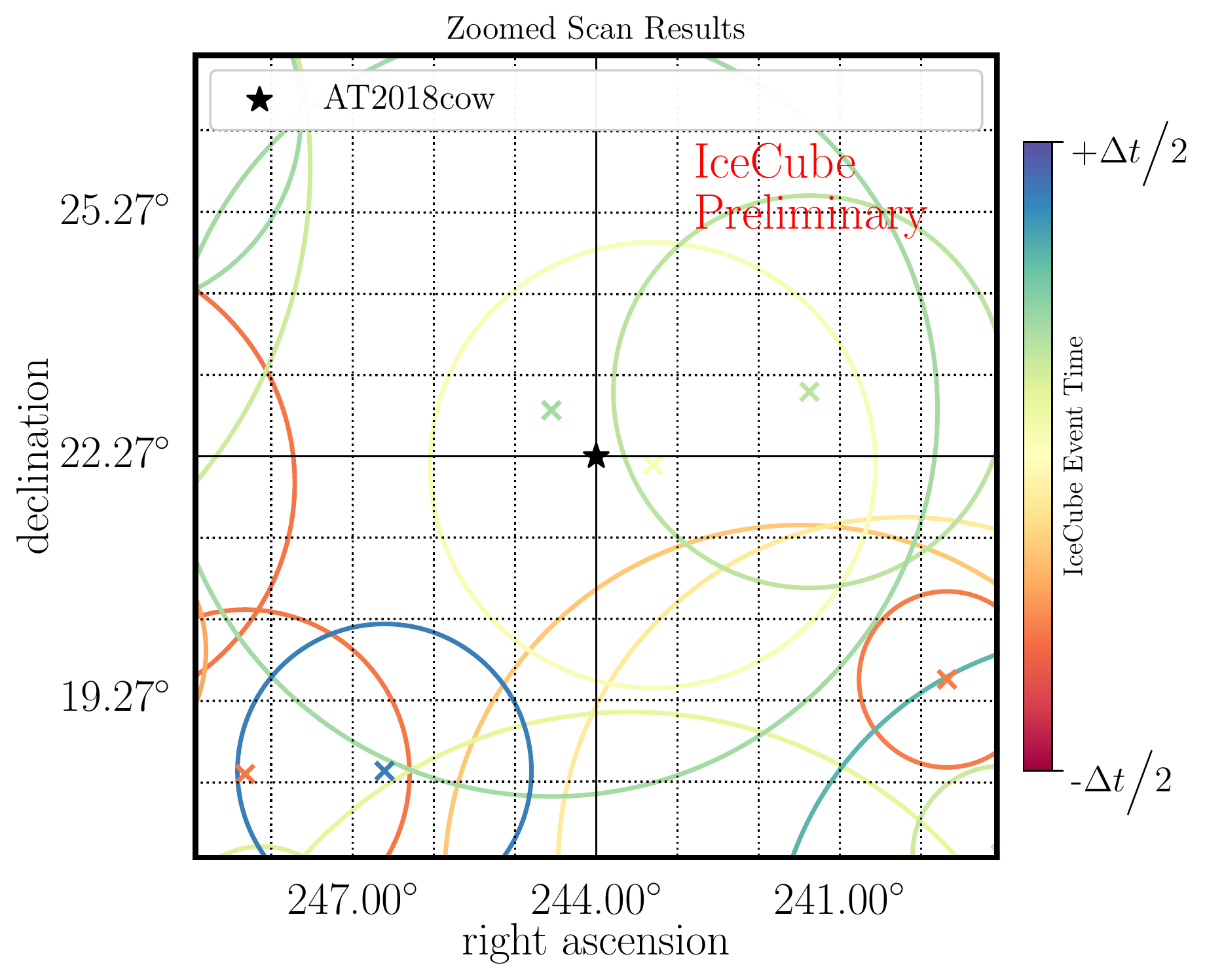}
    \includegraphics[width=2.9in]{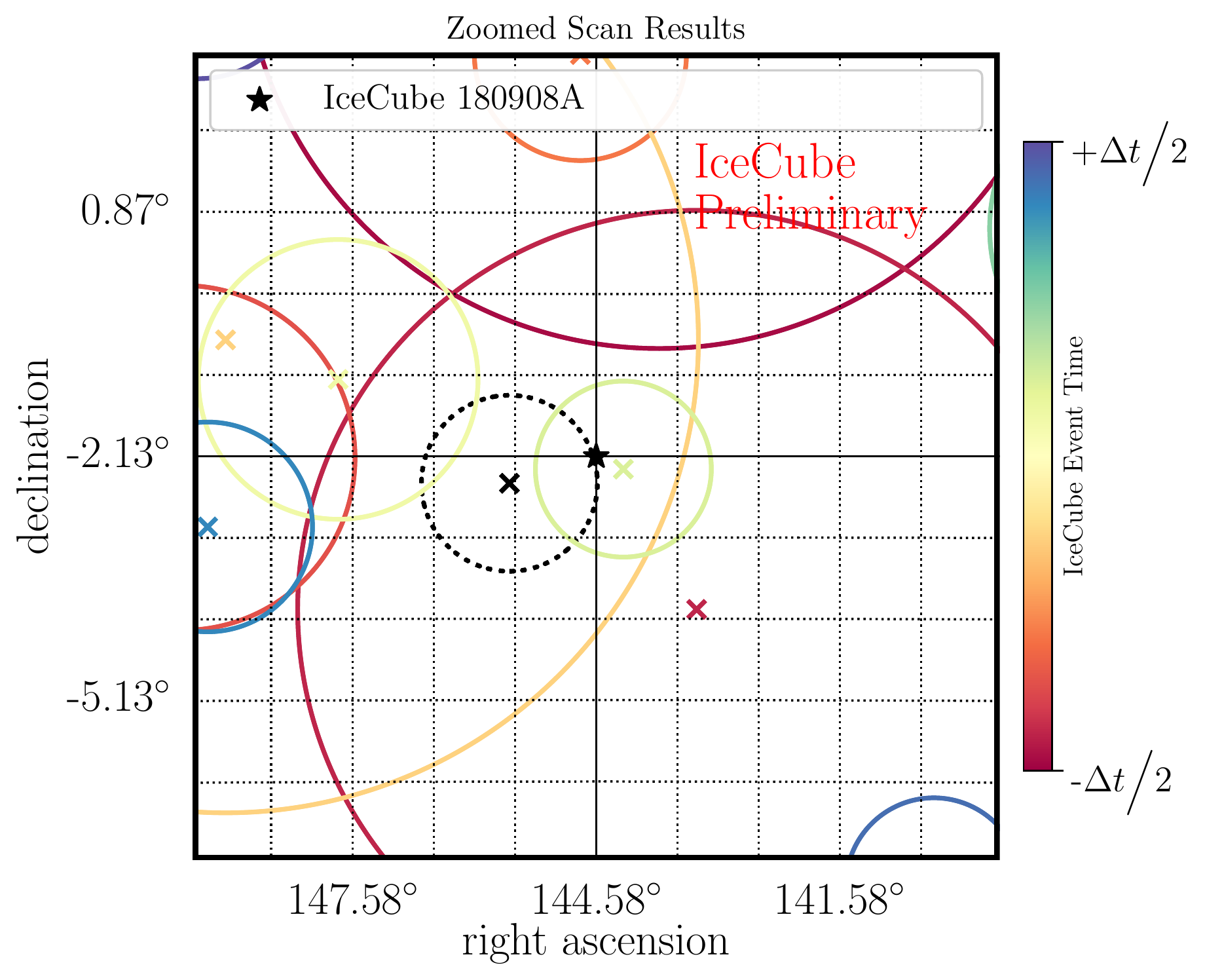}
    \includegraphics[width=2.9in]{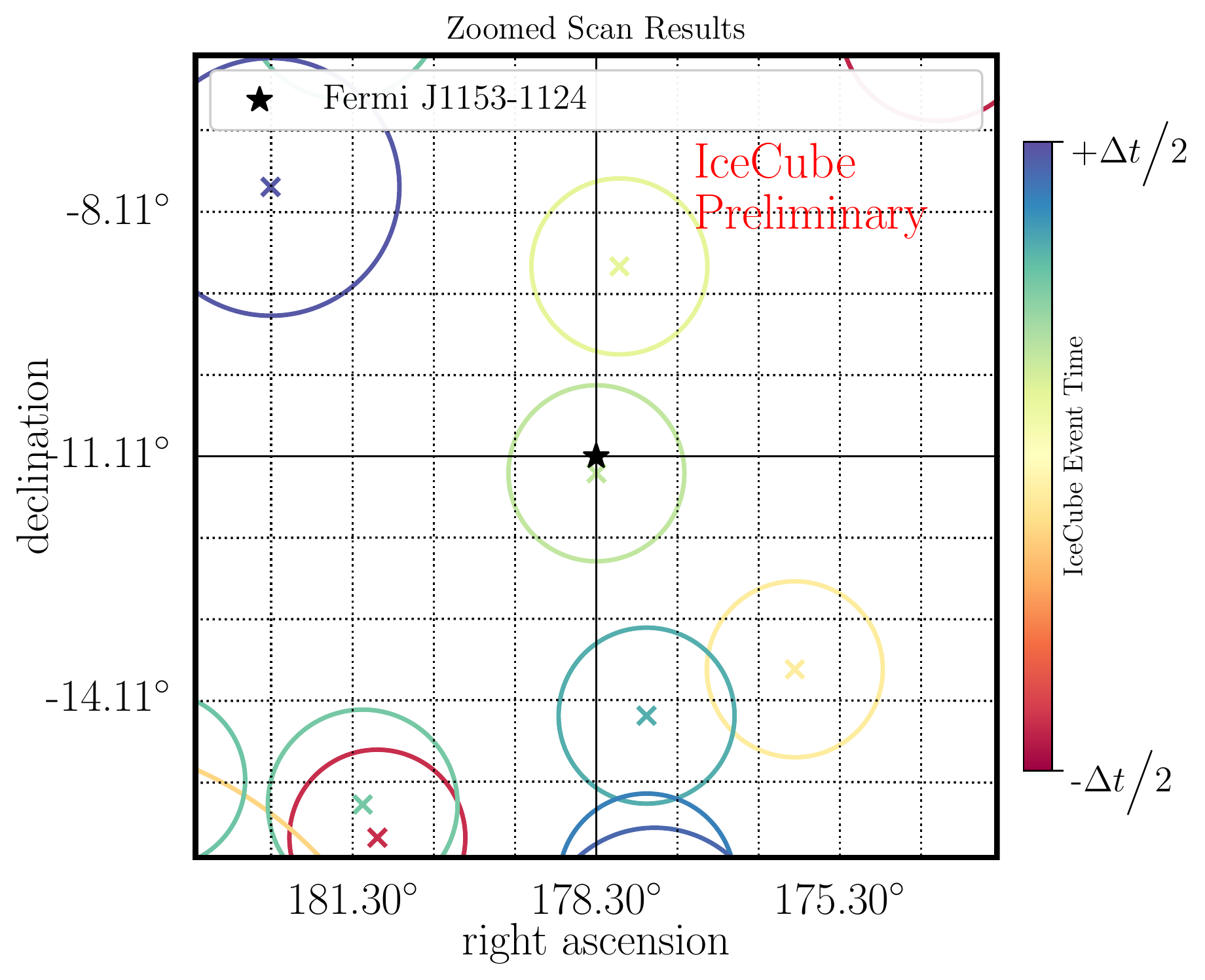}
    \caption{Sky map of IceCube events in the region of interest for the four analyses with smallest p value.  The radius of each event circle indicates its angular uncertainty (90\% containment), and the color indicates its time relative to the center of the search window.  For each analysis, the analysis duration $\Delta$t is given in Table~\ref{table}.  For followup of IceCube's own high-energy event, that event's error circle is indicated with a dashed line and it is not included in the analyzed event sample.\label{skymaps}}
\end{figure}

The four most significant analyses have $p < 0.035$: PKS 0346-27 (a blazar flaring in the GeV band, $p = 0.01$), 2018cow (a mysterious transient with a variety of hypothesized explanations including that it was a bright and unusual supernova or something more exotic, $p = 0.03$), IC-180908A (an IceCube event from the extremely high energy stream, $p = 0.03$), Fermi J1153-1124 (a blazar flaring in the GeV band, $p = 0.02$).  For each of these analyses, Figure~\ref{skymaps} shows a sky map of IceCube events detected during the analysis time window and within the region of interest.

No analysis by the pipeline has yet resulted in a statistically significant detection of a signal.  However, the pipeline is a powerful tool that may enable a future discovery of a neutrino counterpart of a transient/flaring source identified using another messenger or a discovery of a lower-energy neutrino counterpart associated with one of IceCube's high-energy alerts.  Either would provide new information essential for determining the sources and emission mechanisms of astrophysical neutrinos.  Now that the pipeline is executed routinely and is well understood, search results including quantitative upper limits are typically reported publicly via GCN or ATel.  Members of the astronomical community as well as interested members of the public can monitor these channels to determine whether IceCube has analyzed any particular source of interest.  Thanks to IceCube's high duty cycle, full-sky field of view, and low latency, this stream of analysis results reported publicly in near real time may soon include a new discovery.

% 2018gep ($p = 0.04)

%Name &  $\alpha$ ($^{\circ}$) & $\delta$ ($^{\circ}$) & Time (mjd) & Duration (s) &  p-value & Resources 

\begin{longtable}{lrlllrl}
%\toprule

\captionsetup{width=1.00\textwidth}
\toprule
\caption{Summary of all completed fast response analyses as of July 10, 2019.  References are given for those results that have been reported individually in public notices through Gamma-ray Coordinates Network (GCN) or Astronomer's Telegram (ATel).}\label{table}\\

\midrule
Name &  $\alpha$ ($^{\circ}$) & $\delta$ ($^{\circ}$) & Time (mjd) & Duration (s) &  p-value & Reference \\
\midrule 
\endfirsthead
\midrule
Name &  $\alpha$ ($^{\circ}$) & $\delta$ ($^{\circ}$) & Time (mjd) & Duration (s) &  p-value & Reference \\
\midrule
\endhead
\midrule
\multicolumn{7}{r}{{Continued on next page}} \\
\endfoot
\bottomrule
\endlastfoot

 IC-160427A & 240.33 &  +9.86 &  57504.67 &  1.00e+05 & 1.00 &  - \\
 IC-160731A & 214.40 &  +0.19 &  57597.28 &  3.28e+05 & 1.00 &  - \\
 Cygnus X3 & 308.11 &  +40.96 &  57846.00 &  8.64e+04 & 1.00 &  - \\
 GRB 170405A & 219.83 &  -25.24 &  57848.77 &  1.20e+03 & 1.00 &  - \\
 AGL J0523+0646 & 80.86 &  +6.78 &  57858.49 &  4.32e+05 & 1.00 &  - \\
 IC-170506A & 221.80 &  -26.00 &  57879.03 &  8.64e+04 & 1.00 &  - \\
 AT 2017eaw & 308.68 &  +60.19 &  57883.50 &  2.59e+05 & 0.10 &  - \\
 Fermi J1544-0649 & 236.08 &  -6.82 &  57888.00 &  2.74e+05 & 1.00 &  - \\
 Fermi J1544-0649 & 236.08 &  -6.82 &  57891.17 &  9.36e+05 & 1.00 &  - \\
 AXP 4U\_0142+61 & 26.59 &  +61.75 &  57947.95 &  7.20e+03 & 1.00 &  - \\
 GRB 170714A & 34.35 &  +1.99 &  57948.48 &  4.36e+04 & 1.00 &  - \\
 AGL J1412-0522 & 213.00 &  -5.40 &  57970.12 &  1.73e+05 & 1.00 &  - \\
 AT 2017fro & 259.98 &  +41.68 &  57956.00 &  1.21e+06 & 1.00 &  - \\
 G298048 & 197.45 &  -23.38 &  57982.52 &  1.00e+03 & 1.00 &  - \\
 G298048 & 197.45 &  -23.38 &  57982.53 &  1.21e+06 & 1.00 &  - \\
 IC-170922A & 77.43 &  +5.72 &  58017.87 &  1.73e+05 & 1.00 &  - \\
 TXS 0506+056 & 77.36 &  +5.69 &  58011.00 &  1.21e+06 & 1.00 &  - \\
 IC-171106A & 340.00 &  +7.40 &  58063.28 &  8.64e+04 & 0.08 &  - \\
 PKS 0131-522 & 23.27 &  -52.00 &  58073.00 &  1.73e+05 & 0.06 &  - \\
 GRB 171205A & 167.41 &  -12.59 &  58092.26 &  7.20e+03 & 1.00 &  - \\
 Mrk 421 & 166.11 &  +38.21 &  58106.00 &  1.73e+05 & 1.00 &  - \\
 Mrk 421 & 166.11 &  +38.21 &  58130.00 &  8.64e+05 & 1.00 &  - \\
 HESS J0632+057 & 98.25 &  +5.80 &  58135.00 &  6.05e+05 & 1.00 &  - \\
 TXS 0506+056 & 77.36 &  +5.69 &  58186.00 &  5.53e+05 & 1.00 &  - \\
 CXOU J164710.2-455216 & 251.79 &  -45.87 &  58154.77 &  6.88e+04 & 1.00 &  - \\
 Sgr A* & 266.42 &  -29.01 &  58166.02 &  1.80e+03 & 1.00 &  - \\
3C 279 & 194.05 &  -5.79 &  58223.00 &  3.02e+05 & 1.00 &  - \\
% 3C 279 is FSRQ
 PKS 0346-27 & 57.16 &  -27.82 &  58249.00 &  4.18e+05 & 0.01 &  - \\
 PKS 0903-57 & 136.22 &  -57.58 &  58250.00 &  3.31e+05 & 1.00 &  - \\
 AT 2018cow & 244.00 &  +22.27 &  58282.00 &  2.97e+05 & 0.03 &  \href{http://www.astronomerstelegram.org/?read=11785}{ATel 11785} \\
 2FHL J1037.6+5710 & 159.41 &  +57.17 &  58298.92 &  1.69e+05 & 0.11 &  - \\
 NVSS J163547+362930 & 248.95 &  +36.49 &  58305.50 &  3.46e+05 & 1.00 &  - \\
 FRB 180725A & 6.22 &  +67.05 &  58324.25 &  8.64e+04 & 1.00 &  - \\
 GRB 180728A & 253.57 &  -54.03 &  58327.69 &  7.20e+03 & 1.00 &  - \\
 4C +38.41 & 248.82 &  +38.41 &  58362.38 &  2.59e+05 & 1.00 &  - \\
 % 4C +38.41 is FSRQ
 IGR J17591-2342 & 269.79 &  -23.71 &  58340.50 &  1.50e+06 & 1.00 &  - \\
 HAWC Flare & 101.82 &  +37.61 &  58363.47 &  1.95e+05 & 1.00 &  - \\
 IC-180908A & 144.58 &  -2.13 &  58368.83 &  1.73e+05 & 0.03 &  \href{https://gcn.gsfc.nasa.gov/gcn3/23220.gcn3}{GCN 23220} \\
 GRB 180914A & 52.74 &  -5.26 &  58375.48 &  7.20e+03 & 1.00 &  - \\
 GRB 180914B & 332.45 &  +24.88 &  58375.77 &  4.80e+02 & 1.00 &  - \\
 AT 2018gep & 250.95 &  +41.05 &  58369.17 &  1.42e+06 & 0.04 &  \href{http://www.astronomerstelegram.org/?read=12030}{ATel 12030} \\
 SDSS J00289 & 7.12 &  +20.00 &  58394.50 &  3.02e+05 & 1.00 &  - \\
 % SDSS J00289 is FSRQ
 Crab & 83.63 &  +22.01 &  58391.00 &  1.03e+06 & 1.00 &  - \\
 IC-181014A  & 225.15 &  -34.80 &  58375.49 &  2.68e+06 & 1.00 &  \href{https://gcn.gsfc.nasa.gov/gcn3/23340.gcn3}{GCN 23340} \\
 IC-181014A  & 225.15 &  -34.80 &  58404.49 &  1.73e+05 & 1.00 &  \href{https://gcn.gsfc.nasa.gov/gcn3/23340.gcn3}{GCN 23340} \\
 IC-181023A  & 270.18 &  -8.57 &  58384.69 &  2.68e+06 & 1.00 &  \href{https://gcn.gsfc.nasa.gov/gcn3/23380.gcn3}{GCN 23380} \\
 IC-181023A  & 270.18 &  -8.57 &  58413.69 &  1.73e+05 & 1.00 &  \href{https://gcn.gsfc.nasa.gov/gcn3/23380.gcn3}{GCN 23380} \\
 Fermi J1153-1124 & 178.30 &  -11.11 &  58432.00 &  1.73e+05 & 0.02 &  \href{http://www.astronomerstelegram.org/?read=12210}{ATel 12210} \\
 TXS 0506 +056 & 77.35 &  +5.70 &  58449.00 &  6.05e+05 & 1.00 &  \href{http://www.astronomerstelegram.org/?read=12267}{ATel 12267} \\
 IC-190104A  & 357.98 &  -26.65 &  58486.36 &  1.73e+05 & 1.00 &  \href{https://gcn.gsfc.nasa.gov/gcn3/23613.gcn3}{GCN 23613} \\
 IC-190104A  & 357.98 &  -26.65 &  58456.36 &  2.76e+06 & 1.00 &  \href{https://gcn.gsfc.nasa.gov/gcn3/23613.gcn3}{GCN 23613} \\
 GRB 190114C & 54.51 &  -26.94 &  58497.87 &  3.78e+03 & 1.00 &  \href{http://www.astronomerstelegram.org/?read=12395}{ATel 12395} \\
 IC-190124A  & 307.40 &  -32.18 &  58506.16 &  1.73e+05 & 0.07 &  \href{https://gcn.gsfc.nasa.gov/gcn3/23794.gcn3}{GCN 23794} \\
 IC-190124A  & 307.40 &  -32.18 &  58476.16 &  2.76e+06 & 0.20 &  \href{https://gcn.gsfc.nasa.gov/gcn3/23794.gcn3}{GCN 23794} \\
 IC-190221A  & 268.81 &  -17.04 &  58534.35 &  1.73e+05 & 0.08 &  \href{https://gcn.gsfc.nasa.gov/gcn3/23926.gcn3}{GCN 23926} \\
 IC-190221A  & 268.81 &  -17.04 &  58504.35 &  2.76e+06 & 1.00 &  \href{https://gcn.gsfc.nasa.gov/gcn3/23926.gcn3}{GCN 23926} \\
 IC-190331A  & 337.68 &  -20.70 &  58572.29 &  1.73e+05 & 1.00 &  \href{https://gcn.gsfc.nasa.gov/gcn3/24039.gcn3}{GCN 24039} \\
 IC-190331A  & 337.68 &  -20.70 &  58543.29 &  2.68e+06 & 1.00 &  \href{https://gcn.gsfc.nasa.gov/gcn3/24039.gcn3}{GCN 24039} \\
 Mrk 421 & 166.08 &  +38.19 &  58581.00 &  1.43e+06 & 1.00 &  - \\
 IC-190503A  & 120.28 &  +6.35 &  58605.72 &  1.73e+05 & 1.00 &  \href{https://gcn.gsfc.nasa.gov/gcn3/24409.gcn3}{GCN 24409} \\
 IC-190503A  & 120.28 &  +6.35 &  58576.72 &  2.68e+06 & 1.00 &  \href{https://gcn.gsfc.nasa.gov/gcn3/24409.gcn3}{GCN 24409} \\
 IC-190504A  & 65.79 &  -37.44 &  58606.77 &  1.73e+05 & 0.06 &  \href{https://gcn.gsfc.nasa.gov/gcn3/24410.gcn3}{GCN 24410} \\
 IC-190504A  & 65.79 &  -37.44 &  58577.77 &  2.68e+06 & 1.00 &  \href{https://gcn.gsfc.nasa.gov/gcn3/24410.gcn3}{GCN 24410} \\
 IC-190619A & 343.26 &  +10.73 &  58652.55 &  1.73e+05 & 1.00 &  \href{https://gcn.gsfc.nasa.gov/gcn3/24865.gcn3}{GCN 24865} \\
 IC-190619A & 343.26 &  +10.73 &  58622.55 &  2.76e+06 & 1.00 &  \href{https://gcn.gsfc.nasa.gov/gcn3/24865.gcn3}{GCN 24865} \\
 IC-190704A & 161.85 &  +27.11 &  58667.78 &  1.73e+05 & 1.00 &   \href{https://gcn.gsfc.nasa.gov/gcn3/24988.gcn3}{GCN 24988} \\
 IC-190704A & 161.85 &  +27.11 &  58638.78 &  2.68e+06 & 1.00 &   \href{https://gcn.gsfc.nasa.gov/gcn3/24988.gcn3}{GCN 24988} \\
%\midrule
\end{longtable}

\bibliographystyle{ICRC}
\bibliography{references}

\end{document}